\documentclass[twocolumn,showpacs,showkeys,prc,aps]{revtex4}
\usepackage{graphicx}

\newcommand{ \be }{\begin{equation}}
\newcommand{ \ee }{\end{equation}}
\newcommand{ \bea }{\begin{eqnarray}}
\newcommand{ \eea }{\end{eqnarray}}
\newcommand{ \la }{\langle}
\newcommand{ \ra }{\rangle}

\begin{document}

\title{Causal Diffusion and the Survival of Charge Fluctuations in Nuclear Collisions}
\author{Mohamed Abdel Aziz}
\author{Sean Gavin}   
\affiliation{Physics and Astronomy Department, Wayne State
University, 666 W Hancock, Detroit, MI 48201}
\date{\today}

\begin{abstract}
Diffusion may obliterate fluctuation signals of the QCD phase
transition in nuclear collisions at SPS and RHIC energies. We
propose a hyperbolic diffusion equation to study the dissipation
of net charge fluctuations. This equation is needed in a
relativistic context, because the classic parabolic diffusion
equation violates causality. We find that causality substantially
limits the extent to which diffusion can dissipate these
fluctuations.
\end{abstract}


\maketitle

\section{Introduction}\label{sec:intro}

Net-charge fluctuations are measured in nuclear collisions by many
RHIC and SPS experiments \cite{Mitchell}. Conserved quantities
such as net electric charge, baryon number, and strangeness can
fluctuate when measured in limited rapidity intervals. These
fluctuations occur mainly because the number of produced particles
varies with each collision event due to differences in impact
parameter, energy deposition, and baryon stopping. A variety of
interesting dynamic effects can also contribute to these
fluctuations \cite{EarlyFluctuations}. In particular, fluctuations
of mean $p_t$, net charge, and baryon number may probe the
hadronization mechanism of the quark-gluon plasma
\cite{RajagopalShuryakStephanov}.

Fluctuations of conserved quantities are perhaps the best probes
of hadronization, because conservation laws limit the dissipation
they suffer after hadronization has occurred
\cite{JeonKoch,BowerGavin}. This dissipation occurs by diffusion.
While the effect of diffusion on charge and baryon fluctuations
has been studied in refs.~\cite{ShuryakStephanov,BowerGavin}, the
classic diffusion equations used are problematic in a relativistic
context, because they allow signals to propagate with infinite
speed, violating causality.

In this paper we study the dissipation of net charge fluctuations
in RHIC collisions using a causal diffusion equation. We find that
causality inhibits dissipation, so that extraordinary
fluctuations, if present, may survive to be detected. In
sec.~\ref{sec:diffusion} we discuss how the classic approach to
diffusion violates causality, and present a causal equation that
resolves this problem. The derivation we include facilitates our
work in later sections.  We generalize this equation for
relativistic fluids in nuclear collisions in sec.~\ref{sec:ions}.
Related causal fluid equations for viscosity and heat conduction
have been introduced in \cite{muronga} and \cite{teaney} with
different heavy-ion applications in mind.

Our causal formulation can be crucial for the description of
net-charge fluctuations, which involve rapid changes in the
inhomogeneous collision environment \cite{aziz}. We turn to this
problem in secs.~\ref{sec:fluct}--\ref{sec:discussion}. In
sec.~\ref{sec:fluct} we discuss fluctuation measurements and their
relation to two-particle correlation functions. In
sec.~\ref{sec:fluctDiff} we introduce techniques from
\cite{VanKampen} to compute the effect of causal and classic
diffusion on these correlation functions. We then estimate the
impact of causal diffusion on fluctuation measurements in
sec.~\ref{sec:discussion}. These three sections are close in
spirit to work by Shuryak and Stephanov, where a classic diffusion
model is used \cite{ShuryakStephanov}.

\section{causal diffusion}\label{sec:diffusion}

To understand why the propagation speed in classic diffusion is
essentially infinite, recall that the diffusion of particles
through a medium is equivalent to a random walk in the continuum
limit. The variance of the particle's displacement increases by
$d^2 \equiv \langle\Delta x^2\rangle \propto \Delta t$ in the time
interval $\Delta t$ between random steps. The average propagation
speed $v \sim d/\Delta t$ diverges in the continuum limit, where
$\Delta t\rightarrow 0$ with the diffusion coefficient
$D=d^2/\Delta t$ held fixed. Correspondingly, a delta function
density spike is instantaneously spread by diffusion into a
Gaussian, with tails that extend to infinity.
\begin{figure}
\centerline{\includegraphics[width=2.7in]{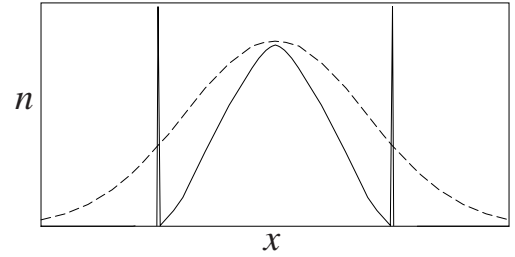}} \vskip
-0.2in \caption[]{Density vs.~position for causal (solid) and
classic (dashed) diffusion assuming an initial delta function
distribution. The spikes represent the right and left moving
diffusion fronts of velocity $v=\pm(D/\tau_d)^{1/2}$.
}\label{fig:causal}\end{figure}

A causal alternative to the diffusion equation is the Telegraph
equation,
\begin{equation}\label{eq:grad}
\tau_d {{\partial^2}\over{\partial t^2}}n +
{{\partial}\over{\partial t}}n = D\nabla^2 n,
\end{equation}
where $D$ is the diffusion coefficient and $\tau_d$ is the
relaxation time for diffusion \cite{telegraph,grad,israel}.
Signals propagate at the finite speed $v=(D/\tau_d)^{1/2}$, so
that a delta function spike spreads behind a front travelling at
$v$, as shown in fig.~\ref{fig:causal}. The classic diffusion
equation omits the term $\propto \tau_d$. Classic diffusion
describes the matter well behind the front at times $t\gg\tau_d$.
This contrasting behavior is generic of hyperbolic equations such
as (\ref{eq:grad}), which include a second order time derivative,
compared to parabolic equations, like the classic first-order
diffusion equation \cite{zauderer}.

Causality concerns are not restricted to relativistic quarks or
pions diffusing through a quark gluon plasma or hadron gas. They
are common -- and constantly debated -- whenever time varying
diffusion and heat conduction phenomena are discussed
\cite{Mandelis}. However, for most nonrelativistic systems,
causality violations are minuscule, so that classic diffusion can
be used. This need not be the case for relativistic fluids
produced in nuclear collisions \cite{muronga,aziz,teaney}.

To motivate (\ref{eq:grad}) and understand its limitations, we
first derive the diffusion coefficient using the Boltzmann
equation in the relaxation time approximation; see, e.g.,
\cite{Gavin85}. Equation (\ref{eq:grad}) can also be obtained by
the moment method \cite{muronga} or by sum-rule arguments
\cite{forster}. For simplicity, we focus on electric charge
transport by a single species. Generalization to multiple species
and other currents such as baryon number is straightforward,
except for the challenging case of color transport
\cite{Mrowczynski}. We describe the evolution of the phase space
distribution $f$ using
\be {{\partial f}\over{\partial
t}}+\mathbf{v}_\mathbf{p}\cdot\nabla f=
-\nu(f-f_e),\label{eq:boltz}\ee
where $\nu^{-1}$ is the relaxation time,
$\mathbf{v}_\mathbf{p}=\mathbf{p}/E$, and $E=\{p^2+m^2\}^{1/2}$.
The local equilibrium distribution satisfies $f_e =
\{\exp[(E-\mu)/T]\pm 1\}^{-1}$ where $T$ is the temperature $T$,
$\mu$ the chemical potential, and the $+$ or $-$ sign describes
bosons or fermions.

Suppose that $\mu$ differs from a local equilibrium value by a
small sinusoidally varying perturbation $\delta\mu$, with $\delta
n= (\partial n/\partial \mu)\delta\mu$. Then $f$ is driven from
local equilibrium by an amount $\delta
f(\omega,\mathbf{k})\exp\{i\mathbf{k}\cdot\mathbf{x}-i\omega t\}$.
Equation (\ref{eq:boltz}) implies
\be\label{eq:domegak0} \delta f(\omega,\mathbf{k}) =
-{i{\mathbf{v}_\mathbf{p}\cdot\mathbf{k}}
\over{\nu-i(\omega-\mathbf{k}\cdot\mathbf{v}_\mathbf{p})}}
{{\partial f_e}\over{\partial\mu}}\delta\mu(\omega,\mathbf{k}),
\ee
plus corrections of order $\nu^{-2}$. The net charge current is
\be \delta \mathbf{j}(\omega,\mathbf{k}) \equiv \int \delta
f(\omega,\mathbf{k}) \mathbf{v}_\mathbf{p} d\mathbf{p}=
-i\mathbf{k} D(\omega,\mathbf{k}) \delta n(\omega,\mathbf{k}), \ee
where $d\mathbf{p}=d^3p/(2\pi)^3$. The diffusion coefficient is
\begin{equation}\label{eq:domegak}
D(\omega,\mathbf{k}) = {{1}\over{3}}{{\partial \mu}\over{\partial
n}} \int
{{v_\mathbf{p}^2}\over{\nu-i(\omega-\mathbf{k}\cdot\mathbf{v}_\mathbf{p})}}
{{\partial f_e}\over{\partial \mu}} d\mathbf{p},
\end{equation}
which is a relativistic generalization of a completely standard
kinetic theory result. For $\mathbf{k}=\omega=0$, we recover the
familiar static diffusion coefficient $D=\nu^{-1}v_{\rm th}^2/3$,
where the thermal velocity $v_{\rm th}=1$ for massless particles.
To obtain quantitative results from this relaxation time
approximation, we identify $\nu^{-1}$ with the relaxation time for
diffusion $\tau_d$ obtained by more sophisticated methods, see
e.g. \cite{Prakash,PVW,HeiselbergPethick,Moore,diffusion}.

To obtain (\ref{eq:grad}), we omit the $\mathbf{k}$ dependence in
(\ref{eq:domegak}), to find
\be D (\omega,0)=D/(1- i\omega \tau_d), \label{eq:diffOmega}\ee
where $D$ is the static diffusion coefficient and
$\nu^{-1}=\tau_d$. We then write
\be (1- i \omega\tau_d)\mathbf{j}(\omega,\mathbf{k})= - i
\mathbf{k}D n(\omega,\mathbf{k}),\label{2}\ee
to find
\be \label{eq:cattaneo}
 \tau_d \frac{\partial} {\partial t}\mathbf{j}(\mathbf{x},t)
+\mathbf{j}(\mathbf{x},t) =- D \mathbf{\nabla} n(\mathbf{x},t),
\label{3} \ee
the Maxwell-Cattaneo relation \cite{telegraph}. Combining
(\ref{3}) with current conservation $\partial n/\partial t +
\mathbf{\nabla}\cdot \mathbf{j} = 0$ yields (\ref{eq:grad}). We
will extend this argument in sec.~\ref{sec:fluctDiff} to derive
eq.~(\ref{eq:VarDiffCaus}).

Classic diffusion follows from (\ref{3}) when the $\tau_d$ term is
negligible, a result known as Fick's law. Fick's law violates
causality because any density change instantaneously causes
current to flow. Including the $\tau_d$ term is the simplest way
to incorporate a causal time lag for this current response
\cite{forster}. Moreover, (\ref{3}) is self consistent in that it
saturates the $f$-sum rule \cite{forster}. However, while
(\ref{eq:grad}) is a plausible approximation, the $k\rightarrow 0$
limit (\ref{2}) is not strictly justified. Possible
generalizations can include the nonlocal equations or gradient
expansions derived from (\ref{eq:domegak0}) and
(\ref{eq:domegak}). If (\ref{eq:grad}) leads to substantial
corrections to classic diffusion, then such generalizations can be
worth considering.

In solving (\ref{eq:grad}) we must impose initial conditions on
the current that respect causality. Suppose that we introduce a
density pulse at $\mathbf{x}=0$ at time $t=0$. Microscopically,
particles begin to stream freely away from this point. The current
implied by (\ref{eq:cattaneo}) is
\begin{equation}\label{eq:catInt}
    \mathbf{j}(t) = \mathbf{j}(0)e^{-t/\tau_d}
    -\int_0^t{{ds}\over{\tau_d}}e^{-(t-s)/\tau_d}D\mathbf{\nabla} n(s).
\end{equation}
Scattering with the surrounding medium eventually establishes a
steady state in which Fick's law holds, as we see from
(\ref{eq:cattaneo}) for $\partial \mathbf{j}/\partial t = 0$, but
this takes a time $t\gg \tau_d$. An assumption of no initial flow
$\mathbf{j}(0) = 0$ is consistent with our physical picture, since
there is no preferred direction for the initial velocity of each
particle. Note that an alternative choice $\mathbf{j}(0) =
-D\mathbf{\nabla} n(0)$ would imply that the current always
follows Fick's law; the corresponding solutions of (\ref{eq:grad})
would never differ appreciably from classic diffusion. However,
this choice is not causal because it requires that particles
``know'' about the medium before they have had any opportunity to
interact with it.

\section{ion collisions}\label{sec:ions}

We now extend (\ref{eq:grad}) to study the diffusion of charge
through the relativistic fluid produced in a nuclear collision.
The fluid flows with four velocity $u^\mu$ determined by solving
the hydrodynamic equations $\partial_\mu T^{\mu\nu}=0$ together
with the appropriate equation of state. We choose the
Landau-Lifshitz definition of $u^\mu$ in terms of momentum current
\cite{degroot}. For sufficiently high energy collisions, the
relative concentration of net charge is small enough that it has
no appreciable impact on $u^{\mu}$. Correspondingly, we take
$u^{\mu}$ to be a fixed function of $\mathbf{x}$ and $t$.

Following \cite{degroot}, we define the co-moving time derivative
and gradient
\be D_\tau\equiv u^\mu\partial_\mu \,\,\,\,{\rm and
}\,\,\,\,\nabla^{\mu}=\partial^{\mu}- u^{\mu}u^{\nu}\partial
_{\nu} \label{eq:coGrad}\ee
for the metric $g^{\mu\nu}={\rm diag}(1,-1,-1,-1)$. In the local
rest frame where $u^\mu = (1,0,0,0)$, these quantities are the
time derivative and $-\mathbf{\nabla}$, where $\mathbf{\nabla}$ is
the three-gradient. The total charge current in the moving fluid
is $j_{\rm tot}^{\mu} = n u^{\mu}+ j^{\mu}$, where the first
contribution is due to flow and the second to diffusion.
Continuity then implies
\be\label{eq:continuity}
\partial_{\mu}(n u^{\mu})=-\partial_{\mu} j^{\mu}. \ee
We assume that the diffusion current satisfies
\be \tau _{d} D_\tau j^{\mu}+j^{\mu}= D \nabla ^{\mu} n
\label{eq:MaxwellCattaneo},\ee
which reduces to (\ref{3}) in the local rest frame.

To illustrate the effect of flow on diffusion, we consider
longitudinal Bjorken flow, $u^\mu=(t/\tau,0,0,z/\tau)$, where
$\tau =(t^2 -z^2)^{1/2}$, $\eta = (1/2) \log ((t+z)/(t-z))$, the
density is a function only of $\tau$ and the current is
$j^{\mu}=(j^t,0,0,j^z)$ \cite{Bjorken}. The continuity equation
(\ref{eq:continuity}) is then
\be \left(\frac {\partial}{\partial \tau}+\frac{1}{\tau}\right)n =
\frac {1}{\tau} \frac {\partial}{\partial \tau} (\tau n) =
-\partial_{\mu} j^{\mu}.\label{eq:contF}\ee
To evaluate the covariant Maxwell-Cattaneo relation, we
differentiate (\ref{eq:MaxwellCattaneo}) to find
\be \tau_{d} \partial_{\mu} (u^{\nu}\partial_{\nu} j^{\mu}) +
\partial_{\mu} j^{\mu} = - D \nabla^{2} n, \label{20}\ee
where $\nabla^2\equiv \nabla_\mu\nabla^\mu$. We then write
\begin{eqnarray}\label{22}
 \partial_{\mu} (u^{\nu} \partial _{\nu} j^{\mu})  &=&
(\partial_{\mu} u^{\nu})(\partial_{\nu} j^{\mu}) + u^{\nu}
\partial_{\nu}(\partial_{\mu} j^{\mu})\nonumber\\
   &=&\frac{1}{\tau} \partial_{\mu} j^{\mu} + \frac {\partial}{\partial
\tau}(\partial_{\mu} j^{\mu}),
\end{eqnarray}
where the second line follows from eqs.~(17) and (20) of
ref.~\cite{Bjorken}. Then (\ref{20}) and (\ref{22}) imply
\begin{equation}\label{eq:MCF}
\tau_d \frac {\partial}{\partial \tau} (\tau\partial_{\mu}
j^{\mu})+\tau\partial_{\mu} j^{\mu} =-D\nabla^2n\tau.
\end{equation}
Together, (\ref{eq:contF}) and (\ref{eq:MCF}) describe causal
diffusion.

To obtain an equation analogous to (\ref{eq:grad}) for the
expanding system, observe that the rapidity density $\rho\equiv
dN/d\eta = A_{\bot} n \tau$, where $A_{\bot}$ is the transverse
area of the two colliding nuclei. If one identifies spatial
rapidity $\eta$ with the momentum-space rapidity of particles,
then $\rho$ is observable. We combine (\ref{eq:contF}) and
(\ref{eq:MCF}) to find that this rapidity density satisfies
\be \tau_{d}\frac{\partial ^{2} \rho}{\partial \tau^{2}} +
\frac{\partial \rho}{\partial\tau} =\frac{D}{\tau^{2}} \frac
{\partial ^{2} \rho}{\partial \eta^{2}}\label{25},\ee
where $\nabla^{2}=\tau^{-2}\partial^2 /\partial\eta^2$ for
longitudinal expansion. Initial conditions for $\rho$ and
$\partial \rho/\partial\tau$ at the formation time $\tau_o$ must
be specified. In view of the causality argument surrounding
(\ref{eq:catInt}), we assume that the initial diffusion current
$j^\mu \equiv 0$, so that (\ref{eq:contF}) implies $\partial
\rho/\partial\tau \equiv 0$ at $\tau=\tau_o$. Note that the total
current $nu^\mu +j^\mu$ is initially non-zero, since the
underlying medium is not at rest.

In the absence of diffusion, longitudinal expansion leaves $\rho$
fixed, as we see from (\ref{eq:contF}) for $j\equiv 0$. Diffusion
tends to broaden the rapidity distribution. To characterize this
broadening, we compute the rapidity width defined by $V \equiv
\la(\eta-\la\eta \ra)^2 \ra = N^{-1}\int \eta^2\rho d\eta$, where
$\la\eta\ra = 0$ and $N = \int\rho d\eta$. We multiply both sides
of (\ref{25}) by $\eta^{2}$ and integrate to find
\be \tau _{d}\frac{\partial ^{2} { V}}{\partial \tau^{2}} +
\frac{\partial {V}}{\partial\tau} =\frac{2 D}{\tau^{2}}.
\label{27}\ee
Observe that classic diffusion follows from (\ref{27}) for
$\tau_{d}=0$ with $D$ fixed. In that case the width increases by
\be\label{30} \Delta V= \frac{2 D}{\tau_o}\left(1-
\frac{\tau_o}{\tau}\right), \quad\quad\quad\quad{\rm classic}\ee
where $\Delta V\equiv{ V}-{ V}(\tau_o)$.
The rapidity width indeed increases, but the competition between
longitudinal expansion and diffusion limits this increase to an
asymptotic value $V_\infty = 2D/\tau_o$.

\begin{figure}
\centerline{\includegraphics[width=3in]{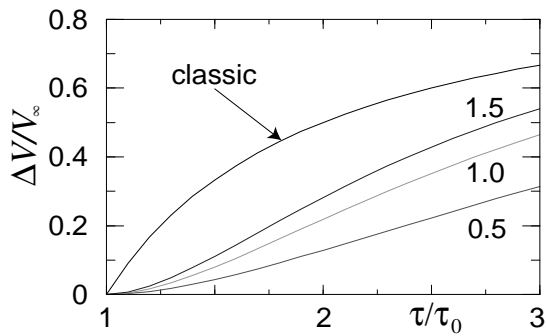}}
\caption[]{Rapidity spread vs. time for causal and classic
diffusion computed using (\ref{28}) and (\ref{30}) respectively.
Causal curves are for $\tau_o/\tau_d = 0.5$, 1, and 1.5. Rapidity
spreads are divided by the asymptotic value $V_\infty =
2D/\tau_o$.}\label{fig:1}\end{figure}

We now solve (\ref{27}) for causal diffusion to obtain
\be\label{28} \Delta V = \frac{2 D}{\tau_o} \int_1^{\tau/\tau_o}
F(\alpha,\lambda) d\lambda,
\ee
where $\alpha \equiv \tau_o/\tau_{d}$ 
and
\be F(\alpha,\lambda)=\alpha\int_1^\lambda \xi^{-2}e ^{\alpha
(\xi-\lambda)} d\xi.\label{29}\ee
We have taken $d V/d\tau = 0$ at $\tau = \tau_o$, as required when
the initial $\partial \rho/\partial\tau$ vanishes.
Figure~\ref{fig:1} compares $\Delta V/V_\infty$ for classic and
causal diffusion. We find that the rapidity broadening is always
slower for causal diffusion compared to the classic case. These
solutions approach one another for $\tau \gg\tau_o$; they are
within 20\% for $\tau > 1.4~\tau_o$ for $\alpha = 10$. The classic
result (\ref{30}) follows from (\ref{28}) and (\ref{29}) for
$\alpha = \tau_o/\tau_d \rightarrow \infty$.

We have so far assumed that $\tau_d$ and $D$ are constant, but
this may not be the case. These coefficients can vary with the
overall density as the system expands and rarefies. As an
alternative extreme, suppose that diffusion occurs as massless
charged particles elastically scatter with the expanding fluid of
density $n_{\rm tot}$. Kinetic theory implies that $\tau_d^{-1}
\approx \langle \sigma v_{\rm rel}\rangle n_{\rm tot}$ and
$D\approx\tau_{d}/3$. If we take the scattering rate $\langle
\sigma v_{\rm rel}\rangle$ averaged over species and temperature to
be constant, but assume $n_{\rm tot}\propto\tau^{-1}$ as
determined by entropy conservation, then $\tau_{d}\propto D\propto
\tau$. Realistically, particle-mass effects would reduce the rate
of growth of $\tau_d$ and $D$, as would the temperature dependence
of $\langle \sigma v_{\rm rel}\rangle$. Nevertheless, the linear
growth rate is worth considering to illustrate how rarefaction can
change the results.

Including this extreme effect of rarefaction implies $\tau_d
=\tau_d(\tau_o)\tau/\tau_o$, so that
\be\tau_d = \tau/\alpha,\label{eq:taudt}\ee
where we now fix the parameter $\alpha=\tau_o/\tau_d(\tau_o)$ at
the initial time. Taking $D = \tau_d /3$ gives
\begin{equation}\label{eq:Dt}
D = \tau/3\alpha,
\end{equation}
so that the diffusion equation (\ref{27}) becomes
\begin{equation}\label{33}
\tau\frac{d^2{V}}{d \tau^2}+\alpha \frac{d
{V}}{d\tau}=\frac{2}{3\tau}.
\end{equation}
As before, classic diffusion is obtained by omitting the second
derivative term. We find
\be\label{eq:Classict}\Delta V = \frac{2 }{3\alpha}\ln
{{\tau}\over{\tau_o}}. \quad\quad\quad\quad{\rm classic}\ee
We see that the width now increases without bound, which is not
surprising since our time varying parameters (\ref{eq:taudt},
\ref{eq:Dt}) imply that $V_\infty = 2D/\tau_o\propto \tau$.

Informed by the classic case, we write the causal equation as
\begin{equation}\label{eq:classDiffT}
\frac{d^2 {V}}{d\theta^2}+(\alpha-1)\frac{d{
V}}{d\theta}=\frac{2}{3},
\end{equation}
where $\theta=\ln {{\tau}/{\tau_o}}$. When $\alpha=1$, the first
derivative vanishes, so that
\begin{equation}\label{37}
 \Delta V=(\ln \tau/\tau_o)^2/3.
\end{equation}
For $\alpha\ne 1$, we find
\begin{equation}\label{38}
\Delta V=\frac{2}{3 (\alpha-1)} \left\{\ln {{\tau}\over{\tau_o}}
-\frac{1}{\alpha-1}\left[1-{\left({{\tau_o}\over{\tau}}\right)^{\alpha-1}}\right]\right\}.
\end{equation}
For $\alpha > 1$, the increase of $\Delta V$ is always smaller
than classic diffusion (\ref{eq:Classict}). In this regime, we see
that $\Delta  {V}/(\Delta {V})_{\rm classic}\rightarrow
\alpha/(\alpha-1)$ as $\tau\rightarrow \infty$.  The difference of
the ratio from unity in the long time limit reflects the fact that
the rarefaction rate $\tau^{-1}$ and diffusion rate
$\tau_d^{-1}\approx \nu$ are in fixed proportion for all time, due
to (\ref{eq:taudt}).

For $\alpha<1$, i.e. for $\tau_o < \tau_d$, eq.~(\ref{38}) implies
that $\Delta {V}$ increases faster than in classic diffusion for
$\tau\gg \tau_o$. This behavior is an unphysical artifact of the
assumption (\ref{eq:taudt}). In this regime the rarefaction rate
$\tau^{-1}$ exceeds the scattering rate $\tau_d^{-1}\approx \nu$,
so that scattering cannot maintain local thermal equilibrium. Our
hydrodynamic diffusion description is therefore not applicable --
one must turn to transport theory, as discussed in \cite{therm}.
One can see this explicitly by solving (\ref{25}) for
$\rho(\eta,\tau)$. Linear perturbation analysis reveals that the
solutions are unstable for $\alpha < 1$. We emphasize that for
constant coefficients the solutions (\ref{30}) and (\ref{28}) are
physical for all $\alpha$, as generally holds for $\tau_d \propto
\tau^{z}$ for any $z < 1$.

\begin{figure}
\centerline{\includegraphics[width=3in]{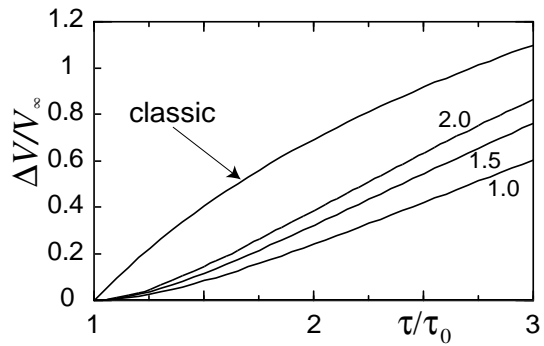}}
\caption[]{Rapidity spread vs. time for causal and classic
diffusion computed using (\ref{37}), (\ref{38}) and
(\ref{eq:Classict}) respectively. Causal curves are for
$\tau_o/\tau_d = 1$, 1.5, and 2. Rapidity spreads are divided by
the value $2D_o/\tau_o=2\alpha/3$.}\label{fig:3}\end{figure}
The width computed with varying coefficients increases without
bound, albeit slowly. This behavior distinguishes (\ref{37}) and
(\ref{38}) from the constant-coefficient widths obtained from
(\ref{30}) and (\ref{28}). Nevertheless, the short time behavior
shown in fig.~\ref{fig:3} resembles the constant-coefficient case
in fig.~\ref{fig:1}. In both cases the classic and causal results
converge over longer times. Note that the results in
fig.~\ref{fig:3} are normalized to $2D/\tau_o = 2/3\alpha$ using
the value of $D$ at $\tau_o$.

To apply these results to collisions, we must specify the
diffusion coefficient $D$ and the relaxation time $\tau_d$.
Transport coefficients in quark gluon plasma and hadron matter
have been been studied extensively
\cite{Gavin85,Prakash,PVW,HeiselbergPethick,Moore,diffusion}.
Flavor and charge diffusion estimates in a plasma yield $D\sim
1-3$~fm and $\tau_d\sim 3D\sim 3 - 9$~fm \cite{HeiselbergPethick}.
The hadron gas case, which is more relevant to this work, is
complicated by the menagerie of resonances produced in collisions.
In ref.~\cite{diffusion}, the charge diffusion coefficient was
computed using a hadronic transport model, yielding $D\approx
2$~fm and $\tau_d\approx 6$~fm.

We argue in sec.~\ref{sec:discussion} that the short time behavior
in figs.~\ref{fig:1} and \ref{fig:3} describes diffusion following
hadronization. Hadrons form at a rather late time, roughly $\tau_o
\sim 6-12$~fm. If freeze out occurs shortly thereafter, at
$\tau_f$ less than 20 fm, then the range $\tau/\tau_o < 3$ is
important. A hadronic relaxation time $\tau_d \sim 6$~fm is
relevant, so that $1 < \alpha < 2$. In this case causal diffusion
gives a much slower spread in rapidity than classic diffusion.

\section{Fluctuations and Correlations}\label{sec:fluct}

Let us now turn to net charge fluctuations and their dissipation.
To begin, we review key features of multiplicity and charge
fluctuation observables. Our discussion builds on the detailed
treatment in ref.~\cite{PruneauGavinVoloshin}, which stresses the
relation of fluctuation observables to two-particle correlation
functions. We extend that treatment by identifying the net-charge
correlation function (\ref{eq:Q}) that drives dynamic charge
fluctuations. In the next section we will show how diffusion
affects the evolution of this correlation function.

Dynamic fluctuations are generally determined from the measured
fluctuations by subtracting the statistical value expected, e.g.,
in equilibrium \cite{PruneauGavinVoloshin}. Dynamic multiplicity
fluctuations are characterized by
\begin{equation}\label{eq:DynamicMult}
    R_{aa}={{\langle N^2\rangle -\langle N\rangle^2 -\langle N\rangle}\over{\langle
    N\rangle^2}},
\end{equation}
where $\langle \cdots\rangle$ here is the event average. This
quantity is obtained from the multiplicity variance by subtracting
its Poisson value $\langle N\rangle$.
Similarly, the covariance for different species is
\begin{equation}\label{eq:DynamicCov}
  R_{ab}= {{\langle N_aN_b\rangle -\langle N_a\rangle\langle N_b\rangle }\over
    {\langle N_a\rangle\langle N_b\rangle}},
\end{equation}
which also vanishes for Poisson statistics. These quantities
depend only on the two-body correlation function
\begin{equation}\label{eq:corrFunExp}
    r_{ab}(\eta_1, \eta_2) = \rho_{ab}(\eta_1,\eta_2)
    - \rho_a(\eta_1)\rho_b(\eta_2),
\end{equation}
where $\rho_{ab}(\eta_1,\eta_2)=dN_{ab}/d\eta_1d\eta_2$ is the
rapidity density of particle pairs for species $a$ and $b$ at the
respective pseudorapidities $\eta_1$ and $\eta_2$, and $\rho_a$ is
the single particle rapidity density. We use the same symbol for
momentum and configuration space rapidity, because we will take
these quantities to be equal in later sections. In
\cite{PruneauGavinVoloshin} it is shown that
\begin{equation}\label{eq:CorrExp}
    R_{ab}={{1}\over{\langle N_a\rangle\langle N_b\rangle}}\int\! d\eta_{1}d\eta_{2}\,
    r_{ab}(\eta_{1},\eta_{2}).
\end{equation}
These quantities are robust in that they are normalized to
minimize the effect of experimental efficiency and acceptance.

The STAR and CERES experiments characterize dynamic net-charge
fluctuations using the robust variance
\begin{equation}\label{eq:nuDyn}
    \nu\equiv R_{++}+R_{--}-2R_{+-},
\end{equation}
proposed in \cite{PruneauGavinVoloshin}; here we use $\nu$ rather
than the more conventional notation $\nu_{dyn}$ for simplicity. To
establish the relation of $\nu$ to the net charge correlation
function
\begin{equation}\label{eq:Q}
    q(\eta_1, \eta_2) = r_{++} + r_{--}
    - r_{+-} - r_{-+},
\end{equation}
we integrate over a rapidity interval to find
\begin{equation}\label{eq:Omega}
 N^2\Omega
 \equiv \int\!\!\!\int q d\eta_1
d\eta_2
   =N(\omega_q - 1),
\end{equation}
where the variance of the net charge is $N\omega_q = {\langle(N_+
- N_-)^2\rangle-\langle N_+ - N_-\rangle^2}$ and the average
number of charged particles is $N \equiv \la N_+ + N_-\ra$.
Expanding (\ref{eq:Omega}) yields
\begin{equation}\label{eq:OmegaR}
 \Omega = f_+^2R_{++} +f_-^2R_{--} -2f_+f_-R_{+-},
\end{equation}
where $f_+ = \la N_+\ra/N = 1-f_-$. If the average numbers of
particles and antiparticles are nearly equal,
\begin{equation}\label{eq:nuOmega}
 \nu\approx 4\Omega.
\end{equation}
HIJING simulations show that this relation is essentially exact
for mesons at SPS and RHIC energy (we find $\nu$ and $4\Omega$ are
equal with 2\% statistical error at SPS energy). We mention that
$\Omega$ itself was suggested as an observable in
ref.~\cite{BowerGavin} and that PHENIX measures the related
quantity $\omega_Q = 1 + N\Omega$. However, $\Omega$ is not
strictly robust. The next section implies that $q$ and,
consequently, $\Omega$ is of more fundamental interest than $\nu$.
However, in view of (\ref{eq:nuOmega}) and the qualitative aims of
fluctuation studies, we will regard these quantities as
interchangeable.

\section{Correlations and Diffusion}\label{sec:fluctDiff}

To understand how diffusion can dissipate dynamic fluctuations of
the net charge and other conserved quantities, we apply the
theoretical framework developed by Van Kampen and others
\cite{VanKampen}. A relativistic extension of these techniques
will enable the computation of the correlation function
(\ref{eq:Q}) as well as statistical quantities like $\nu$, while
introducing no additional parameters.

We start with a single charged species of conserved density
$n(\mathbf{x},t)$ that evolves by diffusion. Consider an ensemble
of events in which $n(\mathbf{x},t)$ is produced with different
values at each point with probability $P\{n(\mathbf{x}, t)\}$.
After ref.~\cite{VanKampen}, one writes a master equation
describing the rate of change of $P$ in terms of transition
probabilities to states of differing $n$ on a discrete spatial
lattice. Diffusion and flow are described as transitions in which
particles ``hop'' to and from neighboring points. The
Fokker-Planck formulation in ref.~\cite{ShuryakStephanov} can be
obtained from this master equation in the appropriate limit.
Alternatively, one can use the master equation to obtain a partial
differential equation for the density correlation function by
taking moments of $P$. The derivation is standard and we do not
reproduce it here \cite{VanKampen}.
%

If the one body density follows classic diffusion dynamics, then
the density correlation function
\begin{equation}\label{eq:Variance}
    r(\mathbf{x}_1,\mathbf{x}_2) \equiv\langle n_1n_2\rangle - \langle
    n_1\rangle\langle n_2\rangle - \delta(\mathbf{x}_1-\mathbf{x}_2)\langle
    n_1\rangle,
\end{equation}
satisfies the classic diffusion equation
\begin{equation}\label{eq:VarDiff}
    \left({{\partial}\over{\partial t}} -
    D(\nabla_1^2 + \nabla_2^2)\right)r(\mathbf{x}_1,\mathbf{x}_2) =
    0
\end{equation}
\cite{VanKampen}. This density correlation function is analogous
to rapidity-density correlation function (\ref{eq:corrFunExp}). If
the event-averaged single-particle density satisfies
(\ref{eq:grad}), then we can extend this result to causal
diffusion by fourier transforming (\ref{eq:VarDiff}) and using
(\ref{eq:diffOmega}), as in sec.~\ref{sec:diffusion}. The inverse
transform yields
\begin{equation}\label{eq:VarDiffCaus}
    \left(\tau_d{{\partial^2}\over{\partial t^2}}+{{\partial}\over{\partial t}} -
    D(\nabla_1^2 + \nabla_2^2)\right)r(\mathbf{x}_1,\mathbf{x}_2) =
    0.
\end{equation}
We remark that the delta-function term in (\ref{eq:Variance})
implies that the volume integral of $r$ vanishes when particle
number fluctuations obey Poisson statistics.  This term is derived
along with (\ref{eq:VarDiff}) in ref.~\cite{VanKampen}; it ensures
that correlations vanish in global equilibrium.

In a nuclear collision with several charged species, only the net
charge density $n^+ - n^-$ satisfies diffusion dynamics, since
chemical reactions can change one species into another. We define
the net charge correlation function $q(\mathbf{x}_1,\mathbf{x}_2)$
as
\begin{eqnarray}\label{eq:cgflc0}
  q&\equiv&
\la (n_1^+ - n_1^-)(n_2^+ - n_2^-)\ra-\la n_1^+ - n_1^-\ra\la
n_2^+ - n_2^-\ra
   \nonumber\\
   & &-\delta(\mathbf{x}_1-\mathbf{x}_2)\la n_1^+ +n_1^-\ra.
\end{eqnarray}
Observe that the integral
\begin{eqnarray}\label{eq:cgflc}
    \int q d\mathbf{x}_1d\mathbf{x}_2 &=& \la (N_+-N_-)^2\ra - \la
    N_+-N_-\ra^2 \nonumber\\
    & &-\la N_++N_-\ra
\end{eqnarray}
gives the net charge fluctuations minus its value for uncorrelated
Poisson statistics. This integral vanishes in equilibrium. We
expand (\ref{eq:cgflc0}) to write $q = r_{++} + r_{--} - r_{+-} -
r_{-+}$, where each $r_{++}$ and $r_{--}$ has the form
(\ref{eq:Variance}) and $r_{+-} \equiv\langle n_1^+ n_2^-\rangle -
\langle n_1^+\rangle\langle n_2^-\rangle$.

To obtain evolution equations for the matter produced in ion
collisions, we incorporate longitudinal expansion following
sec.~\ref{sec:ions}.  The net-charge correlation function
$q(\eta_1, \eta_2,\tau)$ is given by (\ref{eq:Q}), where we now
take
\begin{equation}\label{eq:corrRap}
r_{ab} \equiv \la\rho_1^a\rho_2^b\ra -\la\rho_1^a\ra\la\rho_2^b\ra
-\delta_{ab}\delta(\eta_1-\eta_2)\la\rho_1^a\ra.
\end{equation}
This correlation function has the same form as (\ref{eq:cgflc0})
with densities $n$ replaced by rapidity densities $\rho = dN/d\eta
\propto n\tau$ and $\mathbf{x}_{1,2}$ replaced by spatial
rapidities $\eta_{1,2}$. We find that $q(\eta_1, \eta_2,\tau)$
obeys
\bea\label{40}
 \left(\tau_d \frac{\partial^2}{\partial \tau^2}+
\frac{\partial}{\partial \tau}  -D( \nabla_1 ^2 + \nabla_2 ^2
)\right)q=0  \eea
for $\nabla_{1,2}^2=\tau^{-2}\partial^2/\partial\eta_{1,2}^2$.

To compute the effect of diffusion on net charge fluctuations in
the next section, we write (\ref{40}) in terms of the relative
rapidity $\eta_r \equiv \eta_1 -\eta_2$ and average rapidity
$\eta_a=(\eta_1 + \eta_2)/2$:
\be
 \left(\tau_d \frac{\partial^2}{\partial \tau^2}+
\frac{\partial}{\partial \tau} -
{{2D}\over{\tau^2}}{{\partial^2}\over{\partial\eta_r^2}}
-{{D}\over{\tau^2}}{{\partial^2}\over{\partial\eta_a^2}} \right) q
= 0; \label{41} \ee
the ``2'' follows from the transformation to relative rapidity
$\eta_r$. To compute the widths of $q(\eta_r, \eta_a,\tau)$ in
relative or average rapidity, one multiplies (\ref{41}) by
$\eta_r^2$ or $\eta_a^2$ and integrates over both variables. We
find
\begin{equation}\label{eq:widthRel}
    \Delta \la(\eta_r -\la\eta_r\ra)^2\ra = 2\Delta {V}(\tau)
\end{equation}
and
\begin{equation}\label{eq:widthAvg}
    \Delta \la(\eta_a -\la\eta_a\ra)^2\ra =
    \Delta {V}(\tau),
\end{equation}
where $\Delta {V}(\tau)$ is calculated using (\ref{28}) (or
(\ref{38}) if a time-dependent $\tau_d$ and $D$ is more
appropriate).

\section{Survival of Signals}\label{sec:discussion}

In this section we discuss how diffusion can affect our ability to
interpret hadronization signals. As a concrete illustration, we
suppose that collisions form a quark gluon plasma that hadronizes
at a time $\tau_o$, producing anomalous dynamic charge
fluctuations $\nu$. The hadronization model of Jeon and Koch
\cite{JeonKoch} implies that plasma produces a value $N\nu_{\rm
qgp}\approx -3$. In contrast, the value for a hadronic resonance
gas is variously estimated in the range from $N\nu\approx -1$
\cite{JeonKoch} to $-1.7$ \cite{STAR}. Hadronic diffusion from
$\tau_o$ to a freeze out time $\tau_f$ can dissipate these
fluctuations, reducing $|\nu|$.

We ask whether hadronic diffusion can plausibly bring plasma
fluctuations near the hadron gas values. While the estimates of
ref.~\cite{JeonKoch} can be questioned \cite{Bialas}, they will
serve here as benchmarks. In this section we emphasize the impact
of causal diffusion on this problem, employing a simplified
approach tailored to that aim. Phenomenological conclusions
require a more realistic model.

Before proceeding, we emphasize that ref.~\cite{JeonKoch} compares
quark gluon plasma and hadron gas fluctuations as they might
emerge from distinct systems at equivalent chemical potential and
temperature. In contrast, the fluctuations in a hadron gas that
has evolved \emph{from} a plasma initial state are fixed by charge
conservation. Suppose that hadronization in single event produces
a positive charge in one rapidity subinterval with a compensating
negative charge in another. The final hadronic state maintains
that situation -- freezing in the plasma fluctuations -- unless
diffusion can redistribute the charges between those subintervals.

\begin{figure}
\centerline{\includegraphics[width=3.2in]{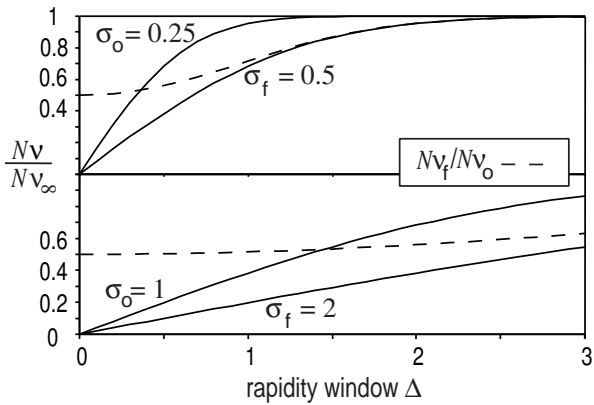}}
\caption[]{Rapidity dependence of dynamic charge fluctuations
assuming diffusion has increased the width from $\sigma_0 = 1$ to
$\sigma_f = 2$ (bottom) and $\sigma_0 = 0.25$ to $\sigma_f = 0.5$
(top).}\label{fig:rapidity}\end{figure}
To compute fluctuation observables following secs.~\ref{sec:ions}
and \ref{sec:fluctDiff}, we specify the correlation function by
identifying spatial and momentum-space rapidity at a fixed freeze
out proper time $\tau_f$. Observe that ISR and FNAL data
\cite{Whitmore} can be characterized as Gaussian near midrapidity.
Moreover, these data show that charged particle correlations are
functions of the relative rapidity $\eta_r=\eta_1-\eta_2$ with
only a weak dependence on the average rapidity
$\eta_a=(\eta_1+\eta_2)/2$. Near midrapidity, these data inspire
the form
\begin{equation}\label{eq:Qyy}
q(\eta_r,\eta_a) \approx {{q_o}\over{2\pi\sigma
\Sigma}}e^{-\eta_r^2/2\sigma^2-\eta_a^2/2\Sigma^2}
\end{equation}
for $\Sigma \gg \sigma$. Diffusion increases the widths $\sigma$
and $\Sigma$ compared to their initial values $\sigma_o$ and
$\Sigma_o$ at the hadronization time $\tau_o$ in accord with
(\ref{eq:widthRel}) and (\ref{eq:widthAvg}). For simplicity, we
assume that $\Sigma_o$ is sufficiently large that we can neglect
the time dependence of $\Sigma$ in (\ref{eq:widthAvg}). We point
out that (\ref{eq:Qyy}) is an exact solution of the classic
diffusion equation on an infinite rapidity interval for Gaussian
initial conditions. Moreover, (\ref{eq:Qyy}) is a good
approximation for causal diffusion, provided that the rapidity
region of interest does not appreciably exceed $\sigma$ or
$\Sigma$; see the discussion in sec.~\ref{sec:diffusion}.

Dynamic fluctuations $\nu$ are computed by integrating $q$ over an
interval $-\Delta/2 \le \eta_1,\, \eta_2 \le \Delta/2$
corresponding to the experimental acceptance. For $\Sigma$ greater
than $\Delta/2$ and $\sigma$, we use (\ref{eq:Omega}) and
(\ref{eq:nuOmega}) to estimate $\nu$
\begin{eqnarray}\label{eq:RapDep}
    N\nu &\approx &
    {{4}\over{N}}
    \int_{-\Delta/2}^{\Delta/2}\int_{-\Delta/2}^{\Delta/2} q(\eta_1,\eta_2)d\eta_1d\eta_2\nonumber\\
    &\approx&
    {{8}\over{N}}\int_0^{\Delta/2}d\eta_a\int_{-\Delta/2+\eta_a}^{\Delta/2-\eta_a}
    q(\eta_r,\eta_a)d\eta_r\nonumber\\
    &\approx& {N\nu}_{{}_\infty}
    \,{\rm erf}\left(\Delta/\sqrt{8}\sigma\right),
\end{eqnarray}
where the total number of charged particles is $N \approx
2\rho\Delta$. We take the rapidity density $\rho$ for each charge
species to be uniform. The quantity $N\nu_{{}_\infty} =
2q_o(2\pi\rho^2\Sigma^2)^{-1/2}$ is the value obtained for a large
rapidity window.

The dependence of the dynamic fluctuations $N\nu$ on the relative
rapidity interval $\Delta$ implied by (\ref{eq:RapDep}) is shown
in fig.~\ref{fig:rapidity} for several ad hoc values of $\sigma$.
Also shown as dashed curves are the ratios of these quantities,
\begin{equation}\label{eq:rat}
{{N\nu_f}\over{N\nu_o}} = {{{\rm
erf}\left(\Delta/\sqrt{8}\sigma_f\right) }\over{{\rm
erf}\left(\Delta/\sqrt{8}\sigma_o\right)}},
\end{equation}
computed for $\sigma_o$ the smaller initial width and $\sigma_f$
the larger final width. This ratio indicates how much the
fluctuations are reduced as the width increases from $\sigma_o$ to
$\sigma_f$. In both figures, the change in widths are chosen so
that they nearly ``hide'' initial QGP fluctuations at the level
expected in ref.~\cite{JeonKoch}, since $N\nu_{\rm hg}/N\nu_{\rm
qgp}\approx 1/3 - 1/2$. Balance function measurements are
consistent with a width $\sigma \approx 0.5$ \cite{Mitchell}. ISR
and FNAL experiments suggest $\sigma \approx 1$ in pp collisions.

\begin{figure}
\centerline{\includegraphics[width=3.2in]{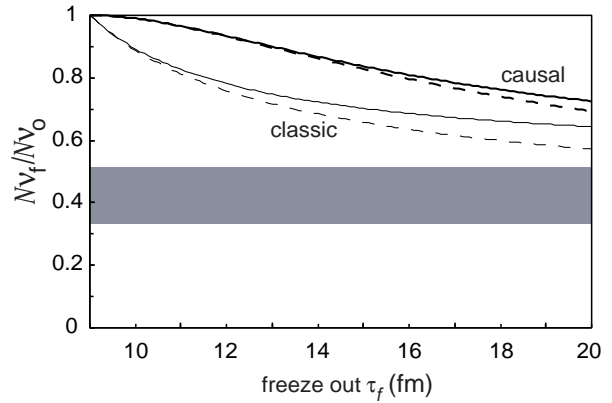}}
\caption[]{Decrease in dynamic charge fluctuations as a function
of freeze out time $\tau_f$ for the acceptance $\Delta = 1$ and
$\sigma_0 = 0.25$. The gray band indicates the level needed to
obscure plasma signals. Solid and dashed curves are respectively
computed for constant and varying coefficients using (\ref{28})
and (\ref{38}) and the corresponding classic eqs.~(\ref{30}) and
(\ref{eq:Classict}).}\label{fig:plasma}\end{figure}
The question then becomes: are such increases in $\sigma$
plausible in a diffusion model? Equation (\ref{eq:widthRel})
implies that $\sigma_f^2 = \sigma_o^2 + 2\Delta {V}(\tau_f)$. The
relevant ``formation time'' is the time at which hadronization
occurs. This occurs quite late in the evolution, roughly from
$\tau_o\sim 6$ to 12~fm. Freeze out occurs later, perhaps as late
as $\tau_f \sim 20$~fm. The ratio $\tau_f/\tau_0$ is then in the
range from one to two, where the difference between causal and
classic diffusion is substantial, as shown in figs.~\ref{fig:1}
and \ref{fig:3}. We take $D\approx 2$~fm and $\tau_d\approx 6$~fm
from ref.~\cite{diffusion}, as discussed in sec.~\ref{sec:ions}.

The solid curves in fig.~\ref{fig:plasma} show the decrease of
dynamic fluctuations computed using (\ref{eq:rat}),
(\ref{eq:widthRel}), and (\ref{28}).  The dashed curves are
computed using (\ref{38}) instead of (\ref{28}). We take the value
$\sigma_o\approx 0.25$ for the initial width; larger values imply
a smaller net reduction of $N\nu$. A hadronization time $\tau_o=
9$~fm is assumed, corresponding to $\alpha = \tau_o/\tau_d = 1.5$
in figs.~\ref{fig:1} and \ref{fig:3}. The gray band indicates the
level to which $N\nu$ must be reduced to hide the effect of plasma
fluctuations. We see that classic diffusion can bring $N\nu$ to
this level if the system lives as long as 20~fm, while causal
diffusion cannot.

\section{Summary}\label{sec:summary}

In this paper we introduce a causal diffusion equation to describe
the relativistic evolution of net charge and other conserved
quantities in nuclear collisions. We find that causal limitations
inhibit dissipation. To study the effect of this dissipation on
fluctuation signals, we obtain a causal diffusion equation for the
two-body correlation function. Our approach complements the
treatment in \cite{ShuryakStephanov}, but is more easily
generalized to include radial flow and other dynamic effects for
comparison to data.

We then use these equations to provide estimates that show that
net charge fluctuations induced by quark gluon plasma
hadronization can plausibly survive diffusion in the hadronic
stage. This result bolsters our optimism for fluctuation and
correlation probes of hadronization and other interesting
dynamics. Our aim here has been to emphasize the impact of causal
diffusion on charge fluctuations. Correspondingly, we employ an
idealized approach tailored to that aim. Phenomenological
conclusions require a more realistic model, to be developed
elsewhere.

Our results on post-hadronization diffusion owe to the similarity
of the lifetime of the hadronic system and the relaxation time for
diffusion $\tau_d$. In fact, RHIC data suggests hadronic lifetimes
that are as short or shorter than we assume: HBT \cite{Lisa} and
resonance \cite{Markert} measurements suggest a lifetime from
roughly $5$ to $10$~fm. On the other hand, our estimate of
$\tau_d$ rests on transport calculations in ref.~\cite{diffusion}.
The importance of this problem invites further study of charge and
other transport coefficients.

In closing, we briefly comment on the experimental situation.
Charge fluctuation measurements give roughly similar values from
SPS to RHIC energy \cite{Mitchell}. Furthermore, the STAR
experiment finds $N\nu\approx -1.5$ at RHIC \cite{STAR}. While
this value is close to the benchmark hadron gas estimates
\cite{JeonKoch,STAR}, we emphasize that those estimates were
constructed for an equilibrium resonance gas assuming no prior
plasma stage. Flow and jet quenching measurements provide strong
indications -- if not proof -- of plasma formation \cite{Miklos}.
As discussed earlier, $N\nu$ can be very different in a resonance
gas formed from a hadronizing partonic system and, indeed, should
be very close to plasma values.

Extraordinary charge fluctuations should be seen, but are not. It
may be that net-charge fluctuations produced by plasma
hadronization are closer to the benchmark hadronic estimates, as
suggested by Bialas \cite{Bialas}. What then? Experimentally, one
can turn to the centrality and azimuthal-angle dependence of
fluctuations and, eventually, to correlation-function measurements
as more sensitive probes of net charge fluctuations. Such studies
have already begun to yield important information
\cite{Mitchell,Gavin04}. In addition, one can study baryon number
and isospin fluctuations, which are more closely related to the
QCD order parameter and, therefore, to hadronization physics and
phase transition dynamics
\cite{RajagopalShuryakStephanov,BowerGavin}.

\section*{Acknowledgements}

We thank R.~Bellwied, H.~Denman, L.~D.~Favro, A.~Muronga,
C.~Pruneau, and D.~Teaney for discussions.  This work was
supported in part by a U.S. National Science foundation CAREER
award under grant PHY-0348559. M.~A.~A. was supported by a Thomas
C. Rumble University Graduate Fellowship at Wayne State
University.


\begin{thebibliography}{99}
%
\bibitem{Mitchell} J.~T.~Mitchell [PHENIX], Proc. Quark Matter 2004,
to be published J.\ Phys.\ G. (2004) arXive:nucl-ex/0404005; G.
Westfall [STAR] \emph{ibid}, arXive:nucl-ex/0404004; C. Pruneau
[STAR], arXiv:nucl-ex/0401016.
%
%
\bibitem{EarlyFluctuations}
  G.~Baym, G.~Friedman and I.~Sarcevic, Phys.\ Lett.\ {\bf B219},
  (1989) 205;
M. Gazdzicki and S.\ Mrowczynski, Z.\ Phys.\ {\bf C54} (1992) 127;
H. Heiselberg, Phys.\ Rept.\ {\bf 351} (2001) 161; S.~Y.~Jeon and
V.~Koch, arXiv:hep-ph/0304012.
%
\bibitem{RajagopalShuryakStephanov}
M.~Stephanov, K.~Rajagopal and E.~V.~Shuryak,
Phys.\ Rev.\ D {\bf 60}, 114028 (1999) [hep-ph/9903292].
%
\bibitem{JeonKoch}
  M.~Asakawa, U.~Heinz, B.~M{\"u}ller, Phys. Rev. Lett. \textbf{85}, 2072
  (2000);
  S.~Jeon and V.~Koch,\emph{ibid}, 2076;
  V.~Koch, M.~Bleicher, S.~Jeon, Nucl.\ Phys.\ \textbf{A698} (2002) 261.
%
\bibitem{BowerGavin} D.~Bower and S.~Gavin, Phys.\ Rev.\ C \textbf{64}
051902 (2001); Y.~Hatta and M.~A.~Stephanov, Phys.\ Rev.\ Lett.\
\textbf{91}, 102003 (2003); Erratum, \emph{ibid} \textbf{91}
129901 (2003).
%
\bibitem{ShuryakStephanov}
E.~V.~Shuryak and M.~A.~Stephanov, Phys.\ Rev.\ C {\bf 63}, 064903
(2001) [arXiv:hep-ph/0010100].
%
\bibitem{muronga}
A.~Muronga, Phys.\ Rev.\ Lett.\ \textbf{88}, 062302 (2002); Phys.
Rev. C \textbf{69}, 034903 (2004).
%
\bibitem{teaney}
D.~Teaney, arXiv:nucl-th/0403053.
%
\bibitem{aziz} M.~Abdel Aziz and S.~Gavin,
Proc. 19th Winter Workshop on Nuclear Dynamics, Breckenridge,
Colorado, USA, February 8-15, 2003, (EP Systema Bt., Debrecen,
Hungary, 2003), 159.
%
\bibitem{VanKampen}
N.\ G.\ Van Kampen, \emph{Stochastic Processes in Physics and
Chemistry}, (Elsevier Science, Amsterdam, 1997);
%
C.~W.~Gardiner, \emph{Handbook of Stochastic Methods for Physics,
Chemistry and the Natural Sciences}, (Springer, Berlin, 2002).
%
\bibitem{telegraph}
C. Cataneo, Atti Semin. Mat. Fis. Univ. Modena \textbf{3}, 3
(1948); P. Vernotte, C. R. Acad. Sci. Paris \textbf{246}, 3154
(1958).
%
\bibitem{grad}
H.~Grad, Comm. Pure Appl. Math. 2, 331 (1949).
%
\bibitem{israel}
W.~Israel and J.~M.~Stewart, Ann.\ Phys.\ (N.Y.) 118, 341 (1979).
%
\bibitem{zauderer} E. Zauderer, \emph{Partial Differential Equations of
Applied Mathematics}, 2nd ed. (Wiley, New York, 1989).
%
\bibitem{Mandelis}
D.~D.~Joseph and L.~Preziosi, Rev.\ Mod.\ Phys.\ \textbf{61}, 41
(1989).
%
\bibitem{Gavin85}
S.~Gavin,
Nucl.\ Phys.\ A {\bf 435}, 826 (1985).
%
\bibitem{forster} D.~M.~Forster, {\em Hydrodynamic fluctuations, Broken
symmetry, and Correlation Functions}, (Benjamin, New York, 1975).
%
\bibitem{Mrowczynski}
C.~Manuel and S.~Mrowczynski, arXiv:hep-ph/0403054.
%
\bibitem{Prakash} M.~Prakash, M.~Prakash, R.~Venugopalan and G. Welke Phys.\ Rept.\ \textbf{227}, 321
(1993).
%
\bibitem{PVW}
A.~Muronga, Phys.\ Rev.\ C \textbf{69}, 044901 (2004).
%
\bibitem{HeiselbergPethick}
H. Heiselberg, C.J. Pethick Phys. Rev. D48, 2916, (1993).
%
\bibitem{Moore}
Peter Arnold, Guy D. Moore, and Lawrence G. Yaffe, JHEP {\bf
0305}, 051, (2003) [arXiv:hep-ph/0302165].
%
\bibitem{diffusion}
N.~Sasaki, O.~Miyamura, S.~Muroya and C.~Nonaka, Europhys.\ Lett.\
{\bf 54}, 38 (2001) [arXiv:hep-ph/0007121].
%
\bibitem{degroot}
S.~R.~de Groot, H.~A.~van Leeuven and C.~G.~van Weert, {\em
Relativistic Kinetic Theory}, (North Holland, Amesterdam, 1980).
%
\bibitem{Bjorken}
J.~D.~Bjorken, Phys.\ Rev.\ D {\bf 27}, 140 (1983).
%
\bibitem{therm}
S.~Gavin,
Nucl.\ Phys.\ B {\bf 351}, 561 (1991).
%
\bibitem{PruneauGavinVoloshin}
C.~Pruneau, S.~Gavin and S.~Voloshin,
Phys.\ Rev.\ C {\bf 66}, 044904 (2002) [arXiv:nucl-ex/0204011].
%
\bibitem{STAR}
J.~Adams et al. [STAR], Phys.\ Rev.\ C {bf 68}, 044905 (2003).
%
\bibitem{Bialas}
A.~Bialas,
Phys.\ Lett.\ B {\bf 532}, 249 (2002) [arXiv:hep-ph/0203047].
%
\bibitem{Whitmore}
J.~Whitmore,
Phys.\ Rept.\  {\bf 27}, 187 (1976); L. Foa, Physics Reports, 22,
1( 1075); H.~Boggild and T.~Ferbel,
Ann.\ Rev.\ Nucl.\ Part.\ Sci.\  {\bf 24}, 451 (l974).
%
\bibitem{BassPratt}
S.~A.~Bass, P.~Danielewicz and S.~Pratt,
Phys.\ Rev.\ Lett.\  {\bf 85}, 2689 (2000)
[arXiv:nucl-th/0005044].
%
\bibitem{Lisa}
M.~Lisa [STAR], Acta Phys.\ Polon.\ {\bf B35}, 37 (2004).
%
\bibitem{Markert}
C.~Markert [STAR], arXiv:nucl-ex/0404003.
%
\bibitem{Miklos}
M.\ Gyulassy, arXiv:nucl-th/0403032.
%
\bibitem{Gavin04} S. Gavin, Phys.~Rev.~Lett., to be
published (2004) [arXiv:nucl-th/0308067]; J.~Phys.~G., to be
published; [arXiv:nucl-th/0404048].

\end{thebibliography}
\end{document}